\documentclass[preprint,aps, preprintnumbers, nofootinbib]{revtex4}
\usepackage[dvips]{graphics}

\begin{document}

\title{Holographic dark energy: quantum correlations against thermodynamical
description}

\author{R. Horvat}
\email{horvat@lei3.irb.hr}
\address{Rudjer Bo\v{s}kovi\'{c} Institute, P.O.B. 180, 10002 Zagreb, Croatia}

\begin{abstract}
Classical and quantum entropic properties of holographic dark energy
(HDE) are considered in view of the fact that its entropy is far
more restrictive than the entropy of a black hole of the same size. In 
cosmological settings (in which HDE is promoted  to a plausible candidate for 
being the dark energy of the universe), HDE should be viewed as  a combined 
state composed of the event horizon and the stuff inside the horizon. By
any interaction of the subsystems, the horizon and the interior
become entangled, raising thereby a possibility that their quantum correlations 
be responsible for the
almost purity of the combined state. Under this circumstances, the
entanglement entropy is almost the same for both subsystems, being also 
of the same  order as the thermal (coarse grained) entropy of the interior 
or the 
horizon. In the context of thermodynamics, however, only additive coarse
grained entropies matter, so we use these entropies to test the generalized
second law (GSL) of gravitational thermodynamics in this framework. While 
we find that the original Li's model passes the GSL test 
for a special choice of parameters, in a 
saturated model with the choice for the IR cutoff in the form of the Hubble
parameter, the GSL always breaks down.

\end{abstract}

\newpage

\maketitle

The concept of holographic principle, first formulated by 't
Hooft \cite{1} and Susskind \cite{2} as a possible window to quantum
gravity, has become part of the mainstream after the Malcadena's discovery 
of AdS/CFT duality \cite{3}. In attempt to reconcile it with the success of
effective-quantum-field-theory description of elementary-particle phenomena, 
the holographic
principle becomes a quantitative measure of the overabundance of degrees of
freedom in ordinary quantum field theory (QFT). Since black holes appear to
involve a vast number of states that are not describable within ordinary QFT,
the entropy for an effective QFT $\sim $ $L^3 {\Lambda }^3 $, where $L$ is
the size of the region
(providing an IR cutoff) and $\Lambda $ is the UV cutoff, should obey the
upper bound \cite{4}
\begin{equation}
L^3 {\Lambda }^3 \leq L^{3/2} M_{Pl}^{3/2} \sim (S_{BH})^{3/4} \ll S_{BH} \;,
\end{equation}
where $S_{BH}$ is the entropy of a black hole of the size $L$.
Since the entropy in QFT scales extensively, it is clear that in an 
expanding universe $\Lambda $ should be promoted to a varying
quantity (some function of $L$ to manifest the UV/IR connection), 
in order (1) not to  be violated during the course of the
expansion. This gives a constraint on the maximum energy density in the
effective theory, $\rho_{\Lambda } \sim {\Lambda }^4 $, to be $\rho_{\Lambda
} \leq  L^{-2}M_{Pl}^2 $. Obviously, $\rho_{\Lambda }$ is the energy density
corresponding to a zero-point energy and the cutoff $\Lambda $. Such a
framework gave rise to a variable cosmological-constant (CC) 
approach generically dubbed
`holographic dark energy' (HDE) \cite{5, 6}, which has proved since to have 
a potential to shed light
both on the `old' CC problem \cite{7} and the `cosmic coincidence problem'
(CCP) \cite{8}. 

The main reason of why the above HDE model is so appealing in possible
description
of dark energy is  when the bound (1) is saturated $\rho_{\Lambda }$ gives
the right amount of  dark energy in the
universe at present, provided $L \simeq H^{-1}$, where $H$ is the Hubble
parameter. Moreover, since $\rho_{\Lambda }$ is a running quantity, it also
has a  potential  to substantially alleviate the CCP. On
the other hand,
the most problematic aspect of the saturated HDE model is its compatibility
with a
transition from decelerated to accelerated expansion. Indeed, as it is well
known, the identification of the IR cutoff with the
Hubble parameter for spatially flat
universes (as suggested by observations) leads to unsatisfactory
cosmologies. In this case one is not able to explain either the accelerating
expansion of the present universe for non-interacting fluids \cite{5}  
or a fact that
the acceleration era has set in just recently, for interacting fluids. A more
realistic class of models, which do allow transitions between the
cosmological eras, is provided by the non-saturated HDE scenario \cite{9, 10}.
As a way out of the above problems, a suggestion of setting $L$ at the future
event horizon has been widely accepted \cite{11}, although inconsistency with
matter dominance irrespective of the choice for $L$ was claimed in 
\cite{10} for any saturated model.

In the present paper, we consider a question of smallness of the upper
bound (1) (with respect to $S_{BH}$) from the aspect of information theory
\cite{12}. Using the formalism and language of the physics of information we 
define fine/coarse grained entropies as well as the entropy of entanglement 
for HDE. Finally, we switch to classical (thermodynamical) description
of the system to test the generalized second law (GSL) of gravitation and
irreversibility for the HDE scenario.

The central question we would like to address here is why the entropy (1) is
so much smaller than the entropy calculated using the
first law of thermodynamics with the temperature of the horizon (the only
temperature we have at our disposal). The latter turns out to be of the order 
of $S_{BH}$ as well (see below). Note that the original model \cite{4} 
leading to (1), aiming to explain the present acceleration of the universe (to
become HDE), leads to cosmological models which do have finite event
horizons. Therefore in cosmological settings, in which the system described
by (1) becomes HDE, we actually deal with two subsystems: the horizon
and the stuff inside the horizon \footnote{We shall deal here only with the CC
stuff inside the horizon since during dark-energy domination its
contribution grossly
overwhelms that of  ordinary matter.}. We will argue that quantum 
mechanical entanglement between the two subsystems can explain the small 
value in (1).

Let us now analyze the situation from the aspect of information theory. The
small entropy in (1) is usually  referred as a fine grained entropy of the 
composite system,
and since it is $\ll S_{BH}$ (as well as $\ll $ than 
other entropies to be defined
below), we will assume, for simplicity, that the composite system is in a
pure state. The results from information theory
\cite{12} then easily apply to our case. The subsystems (the
interior and the horizon), are not generally described by pure states but by 
mixed density matrices, resulting in an entanglement entropy (or fine grained
entropy) for the subsystems, -$Tr\rho_{\Lambda }log \rho_{\Lambda }$ and
-$Tr\rho_{hor}log \rho_{hor}$. It measures both the degree of entanglement
between the subsystems and the departure from a pure state for a particular
subsystem. Furthermore, 
if the initial state of the combined system is pure, the equality of the
entanglement entropies results, $-Tr\rho_{\Lambda }log \rho_{\Lambda } =
-Tr\rho_{hor}log \rho_{hor} \equiv S^{ent}$ \cite{12}. 
In addition, $S^{ent}$ can be also
thought of as a lack of information $I$, defined as $I = S^{ther} -
S^{ent}$, where $S^{ther}$ is the thermal (or coarse grained) entropy, 
representing a distribution which maximizes the entropy for a given 
system at a given average energy. Thus, 
$S^{ther} > S^{ent}$. Note that any thermodynamic considerations involve only 
$S^{ther}$'s. The purity of the combined state and the presence of 
entanglement may result that a great deal of information, $2S^{ent}$, 
to be stored in the
correlations between the subsystems rather than in the subsystems themselves.
Therefore if the information content of the correlations equals $2S^{ent}$,
the correlations between subsystems would make the whole system pure.

Next, let us estimate $S^{ent}$ for the case under consideration. Although 
it is not unambiguously defined because of a lack of knowledge of the system,
the
information theory says \cite{12} that $S^{ent} \simeq S_{BH}$ or
$S^{ther}_{\Lambda }$, depending on the share the subsystems have in the
whole system (see also \cite{13, 14}). $S^{ther}_{\Lambda }$ can be obtained 
using the first law of
thermodynamics with the temperature of the horizon, giving a contribution of
the order of $S_{BH}$ (see below) \footnote{For possible ambiguities see the
footnote on p.5. Thus, either $S^{ther}_{\Lambda }$ or $|S^{ther}_{\Lambda
}|$ is of the order of $S_{BH}$.}. Thus, $S^{ent}$ is typically of
the same order as the horizon entropy.

After having shown qualitatively that quantum correlations between the event
horizon and the interior dark energy given by a HDE variable $\Lambda $
term, may be responsible for a small value (1), we turn to a quantitative 
analysis involving classical (thermodynamical) properties of HDE.
Namely, we put the HDE model under the scrutiny of another
profound physical principle, the GSL of
gravitational thermodynamics. In the context of modern cosmology, the
Second Law of thermodynamics  is manifest there since the initial conditions
for cosmology have low entropy, so we can see the Second Law in operation
\cite{15}. In the problem under consideration it is adequate to invoke the
GSL because we are dealing with cosmologies in which ever accelerating
universes always possess future event horizons. The GSL
states that
the entropy of the event horizon plus the entropy of all the stuff in
the volume inside the horizon cannot decrease in time. The idea of
associating entropy with the horizon area surrounding black holes is
now extended to include all event horizons \cite{16}. 

We aim to restrict the
parameter $c^2$, which helps to parametrize the saturated HDE energy density
$\rho_{\Lambda } = (3/8 \pi)c^2 L^{-2}M_{Pl}^2$ \cite{6}, by assuming the
validity of the GSL. A restriction on $c^2$ under the combined phenomenological
constraints has been recently obtained \cite{17} for certain HDE 
models. For another studies
searching for the conditions required for validity of the GSL in
cosmological models involving dark energy, see
\cite{18}. 

As mentioned
earlier, the GSL involves only thermal entropies which are additive by
definition, and with a macroscopic scale of resolution due to coarse graining,
the arrow of increasing time and irreversibility naturally 
emerge.   The GSL thus states that 
(omitting `ther' from $S^{ther}_{\Lambda }$ hereafter)
\begin{equation}
\dot{S}_{hor} + \dot{S}_{\Lambda } \geq 0 \;.
\end{equation}  
Here overdots represent time derivatives, $S_{hor} = \pi M_{Pl}^2 d_{E}^2$ and
the future event horizon is given by
\begin{equation}
d_{E} = a \int_{a}^{\infty } \frac{da}{a^2 H} \;,
\end{equation}
with $a$ being a scale factor.              

The entropy inside the horizon can be determined using the first law of
thermodynamics
\begin{equation}
T_{\Lambda }dS_{\Lambda } = d(\rho_{\Lambda }V) + p_{\Lambda }dV \;,
\end{equation}
where $T_{\Lambda }$ is the horizon temperature, $V= (4\pi /3) d_{E}^3$ and
$p_{\Lambda } = w_{\Lambda } \rho_{\Lambda }$. We shall examine (4) using
both the event and the apparent horizon in the definition of the temperature
$T_{\Lambda } \equiv 1/(2 \pi d_{E,A})$, where $d_A = H^{-1}$ for flat 
space. Putting all together, the constraint (2) can be written as
\footnote{Actually for the time derivative of $S_{\Lambda }$, we obtain from
(4) that $\dot{S}_{\Lambda } = (1/ 2 T_{\Lambda }) c^2 \dot{L} (1 + 3
w_{\Lambda }) M_{Pl}^2$, showing that for 
$\dot{L} > 0$  $S_{\Lambda }$ starts decreasing at the onset of the accelerated
phase ($w_{\Lambda } \leq -1/3$). 
Taking $T_{\Lambda } \sim L^{-1}$ as usual, we find upon
integration (neglecting an integration constant) that $S_{\Lambda } \sim c^2 L^2
(1 + 3w_{\Lambda }) M_{Pl}^2$, which is obviously negative. To our
knowledge, a situation where $S_{\Lambda }$ is negative even in non-phantom
cosmologies was indicated for the first time in \cite{19}. In this case, the
thermal entropy, which should obviously reflect the number of microscopically 
distinct quantum states, becomes hard to interpret. This also has 
implications for information theory introduced above.} 
\begin{equation}
d_{E,A} \left (-d_{E}^2 L^{-3} \dot{L} + \frac{3}{2} (1 + w_{\Lambda })
L^{-2}
d_{E} \dot{d}_{E}\right ) + \frac{1}{c^2} \dot{d}_{E} \ge 0 \;.
\end{equation}

For the choice $L = d_E $ and using $d_E = c(1 + r)^{1/2} d_A $, obtained
from the Friedmann equation (for flat space) 
with the dominant matter component $\rho_m $ and $r
= \rho_m /\rho_{\Lambda }$, the constraint (5) is reduced further to
\begin{eqnarray}
(d_{A,E}/d_{E}) (1 + 3 w_{\Lambda }) + 2/c^2 \ge 0 \;~\;~\;~ ; \;~\;~\;~ 
\dot{d}_{E} > 0 \;,
\\
(d_{A,E}/d_{E}) (1 + 3 w_{\Lambda }) + 2/c^2 \le 0 \;~\;~\;~ ; \;~\;~\;~
\dot{d}_{E} < 0 \;.
\end{eqnarray}
Let us now test the popular Li's model \cite{6}, with $r = 0$, $w_{\Lambda }
= -1/3 -2/3c $, $d_{E} \sim a^{1 - 1/c}$, against the GSL. One
obtains,
\begin{eqnarray}
-c (d_{A,E}/d_{E}) + 1 \ge 0 \;~\;~\;~ ; \;~\;~\;~ c>1 \;,
\\
-c (d_{A,E}/d_{E}) + 1 \le 0 \;~\;~\;~ ; \;~\;~\;~ c<1 \;.
\end{eqnarray}
Taking first $T_{\Lambda } = 1/(2 \pi d_{E})$, we do obtain a contradiction 
for both constraints (8) and (9) unless $c = 1$; the model
therefore passes the GSL test only for $c^2 =1$. With 
$T_{\Lambda } = 1/(2 \pi d_{A})$ one 
obtains zero on the LHS of either constraint (8-9). This means that total
thermodynamical entropy of the system stays constant during cosmological
evolution in the $\Lambda $-dominated phase. The GSL is therefore 
respected for any $c^2$.

Another plausible choice, $L=H^{-1}$, makes sense only in the presence of
interaction between (near) pressureless dark matter with 
HDE \cite{20, 9}. Otherwise
HDE is not able to bring about an accelerated phase of the present universe
\cite{5}. To obtain a realistic cosmology, a certain degree of 
non-saturation in the HDE energy density is also needed, to result in a
matter-dominated epoch in the past \cite{9, 10}. Since we are going to test
the model under GSL only during accelerated expansion, we shall use the
saturated version of HDE. In this case, the constraint (2) reduces to
\begin{equation}
d_{E,A} \left (d_{E}^2 H \dot{H} + \frac{3}{2} (1 + w_{\Lambda }) H^{2}
d_{E} \dot{d}_{E} \right ) + \frac{1}{c^2} \dot{d}_{E} \ge 0 \;.
\end{equation}         
For a constant interaction parameter, it follows that $\rho_{m},
\rho_{\Lambda } \propto a^{-3m}$ with $m = 1 + c^2 w_{\Lambda }$ \cite{9}.
Also $m < 2/3$, to obtain an accelerated universe. Using this,
all the relevant entries in (10) can easily be obtained. Taking $T_{\Lambda } =
1/(2 \pi d_{A})$, (10) is reduced further to 
\begin{equation}
3 c^2 -1 \geq 2c^2 \;,
\end{equation}
leading to a final constraint $c^2 \geq 1$. However, with the aid of the
Friedmann equation for flat space, one can express $c^2$ for such a choice for
$L$ as 
\begin{equation}
c^2 = \frac{1}{1 + r_0 } \simeq 0.7 \;.
\end{equation}
Hence, the GSL is not respected here. 

Another choice in (10), $T_{\Lambda } = 1/(2 \pi d_{E})$, leads to a bound
\begin{equation}
9 w_{\Lambda }^2 c^4 + (2 + 12 w_{\Lambda }) c^2 + 1 \geq 0 \;,
\end{equation}
which now depends on $w_{\Lambda }$. Observationally, $w_{\Lambda }$ is very
close to $-1$, and $w_{\Lambda } = -1$ is also the most natural value 
for HDE (since in the original derivation it represents zero-point energies).
This means that $c^2$ should reside in the allowable range, $1 < c^2 < 1/9 $.
Since the value (12) obtained from the Friedmann equation does not fit the
above range, we see again that the GSL is not respected. So, the saturated 
HDE model with the choice for the IR cutoff, $L = H^{-1}$, does not respect
the GSL of gravitational thermodynamics.

Let us conclude by laying stress once again on some basic points on which 
this paper resides. We have shown that the entropy for the HDE model as 
given by
(1) should not be used in thermodynamical considerations. Instead, it should
be interpreted as the fine grained entropy of the system composed of the
horizon and the interior dominated by a variable CC term. Stated
differently, even if we assume thermal equilibrium between weakly interacting 
subsystems, the whole system will not be thermal. We have also introduced the entanglement entropy for the
subsystems (their fine grained entropy) to show that, via quantum
correlations, this entropy may be responsible for the (almost) purity of the
entire system. The fine grained entropies for the subsystems are neither
additive nor conserved. On the other hand, any thermodynamical consideration
does involve only thermal (or coarse grained ) entropies, which are additive but,
of course, not conserved. Using these properties we have tested the model
against the GSL, which requires that the thermal entropy of the whole system
(the sum of thermal entropies of the subsystems in thermal equilibrium)
never decreases in the course of cosmic expansion. We have tested
two simplest although distinct models (non-interacting versus interacting),  
to obtain that the model in which the IR cutoff is set by the future event 
horizon, always has a capacity to pass the GSL test.

{\bf Acknowledgment. } This work was supported by the Ministry of Science,
Education and Sport
of the Republic of Croatia under contract No. 098-0982887-2872.

\end{document}